\documentclass[a4paper,12pt]{article}

\usepackage{color}
\usepackage[abs]{overpic}

\usepackage{amssymb}
\usepackage{amsmath,epsfig,subfigure,epstopdf}
\usepackage{graphicx}

\newcommand{\be}{\begin{equation}}
\newcommand{\en}{\end{equation}}

\renewcommand{\vec}[1]{\boldsymbol{#1}}
\newcommand{\ii}{\textrm{i}}
\newcommand{\ee}{\textrm{e}}

\begin{document}

\title{Wrinkles in the opening angle method}

\author{
Michel Destrade$^{a,b}$, Irene Lusetti$^c$, \\
Robert Mangan$^a$, Taisiya Sigaeva$^{d,e}$ \\[12pt]
$^a$School of Mathematics, Statistics and Applied Mathematics, \\ NUI Galway, University Road, Galway, Ireland;\\[6pt]
$^b$School of Mechanical and Materials Engineering, \\
University College Dublin, Belfield, Dublin 4, Ireland; \\[6pt]
$^c$Politecnico di Milano, \\
piazza Leonardo da Vinci 32, 20133 Milano, Italy; \\[6pt]
$^d$Department of Mechanical and Manufacturing Engineering,\\ University of Calgary, Calgary, AB, Canada; \\[6pt]
$^e$Department of Mechanical Engineering,\\ Lassonde School of Engineering, \\ York University, Toronto, ON, Canada.
}
\date{}

\maketitle


\begin{abstract}

We investigate the stability of the deformation modeled by the opening angle method, often used to give a measure of residual stresses in arteries and other biological soft tubular structures. 
Specifically, we study the influence of stiffness contrast, dimensions and  inner pressure  on the onset of wrinkles when an open sector of a soft tube, coated with a stiffer film, is bent into a full cylinder.  
 The tube and its coating are made of isotropic, incompressible, hyperelastic materials.
We provide a full analytical exposition of the governing equations and the associated boundary value problem for the large deformation and for the superimposed small-amplitude wrinkles.
 For illustration, we solve them numerically with a robust algorithm  in the case of Mooney-Rivlin materials.
We confront the results to experimental data  that we collected for  soft silicone sectors.
We study the influence of axial stretch and inner pressure on the stability of closed-up coated tubes with material parameters comparable with those of soft biological tubes such as arteries and veins, although we do not account for anisotropy. 
We find that the large deformation described in the opening angle method does not always exist, as it can become unstable for certain combinations of dimensions and material parameters.

\end{abstract}




\section{Introduction}

\begin{figure}[htbp]
\label{residual}
\centering
\includegraphics[width=1.0\textwidth]{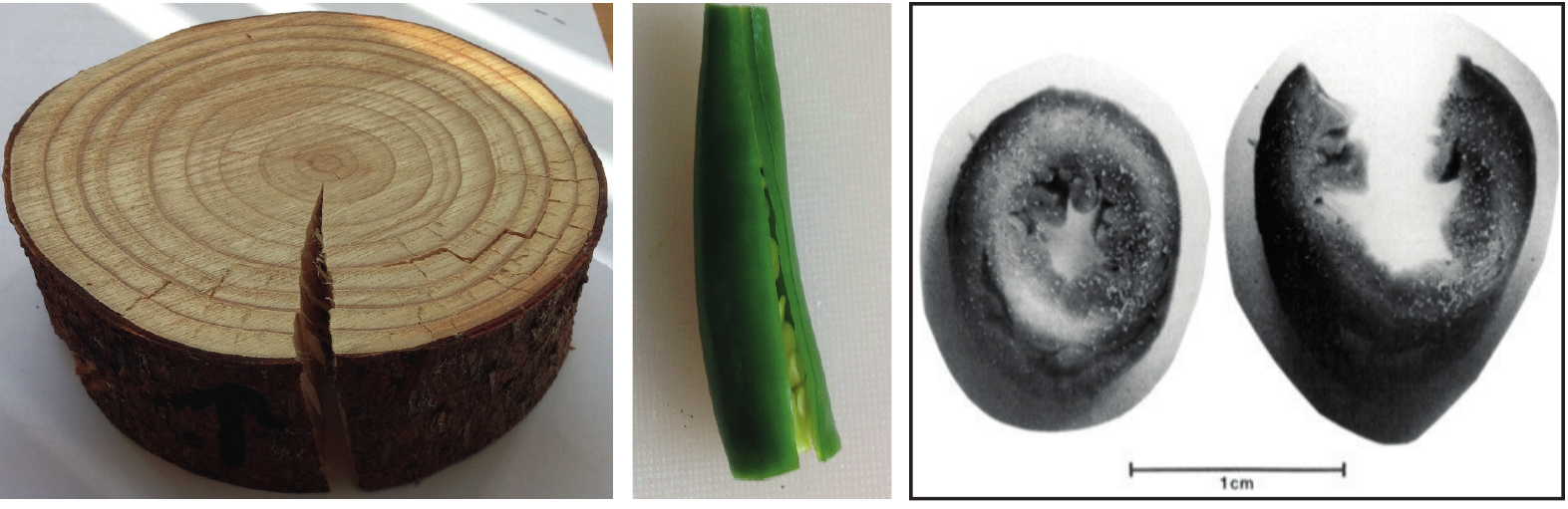}
\caption{Cutting biological cylindrical  structures radially reveals that they were under circumferential residual stresses. Left: slice of an Irish Ash tree;  Middle: a green chilli pepper; Right: Equatorial slice of rat heart (taken from \cite{OmFu90}).}
\end{figure}

One of the most effective ways  to demonstrate the existence of \emph{residual stresses in biological structures} is to isolate a cylindrical shape and cut it axially.
Invariably it will open up, revealing that the cylinder was under a large circumferential stress, see Fig.\ref{residual}.

In turn, one of the most successful advances of non-linear elasticity is the modeling of this stress through the so-called \emph{opening angle method}.
By measuring how much a tube opens up into a sector, one can reconstitute a backward scenario whereby the structure was initially an open circular sector, subsequently bent into a complete tube by the action of what can now be identified as a residual stress, see Fig.\ref{FigConf}.
Hence the opening angle gives a measure of the level of residual stress for an assumed model of material behavior. 

Of course many questions remain open at the end of the process and here we address the following: Is the bending deformation always possible, or is it limited by loss of stability with respect to small-amplitude static wrinkles?
Moreover, can the instability be overcome by pressurization of the reconstituted tube? 
These issues are most relevant to Finite Element simulations of residually-stressed  tubes,  where buckling should be avoided as much as possible.

Here we first formulate in Section \ref{section2} the equations governing the large deformation of a coated circular sector into an intact tube, which is  possibly  subjected to an internal hydrostatic pressure  and a uniform axial stretch.
We then specialize the analysis to the case when the coating and the substrate are  made of different  Mooney-Rivlin materials, because the stress components can then be computed analytically. 
We pay particular attention to writing the boundary conditions properly (hydrostatic pressure on inner face, perfect contact at the interface, traction-free on outer face).

In Section \ref{section3} we present the algorithm implemented to solve the incremental problem of static wrinkles superimposed onto  large bending,  axial stretch,  and pressurizing. 
It relies on the Stroh formulation and the Surface Impedance Matrix method, and is robust and unaffected by numerical stiffness.

Finally, Section \ref{section4} presents experimental and numerical results: first our own, achieved by gluing a silicone coating on a urethane substrate; and second those coming from the literature on  soft biological tubes, although of course those cannot  be accurately modeled as isotropic. In our experiments, we show that no wrinkles form when a sector of opening angle  $120^\circ$ is closed, while wrinkles form before a sector of opening angle  $240^\circ$ is closed. 
Applying the aforementioned algorithm, we show numerically that the critical opening angle at which wrinkles form is  $209^\circ$ and that four wrinkles should appear along the circumference, which is consistent with the experimental results. 
Applying the algorithm for  dimensions and material parameters  comparable (with the limitation that anisotropy is not accounted for) to those  of a rabbit artery, we show that, in the absence of internal pressure, wrinkles form for an opening angle of $320^\circ$,  but that these wrinkles can be eliminated by applying  an internal pressure or can be delayed by the presence of an axial stretch.  These results are in line with intuition  and experiments  made on biological tubes.


\section{The opening angle method}
\label{section2}

\begin{figure}[htbp]
\label{FigConf}
\centering
\includegraphics[width=1.0\linewidth,grid]{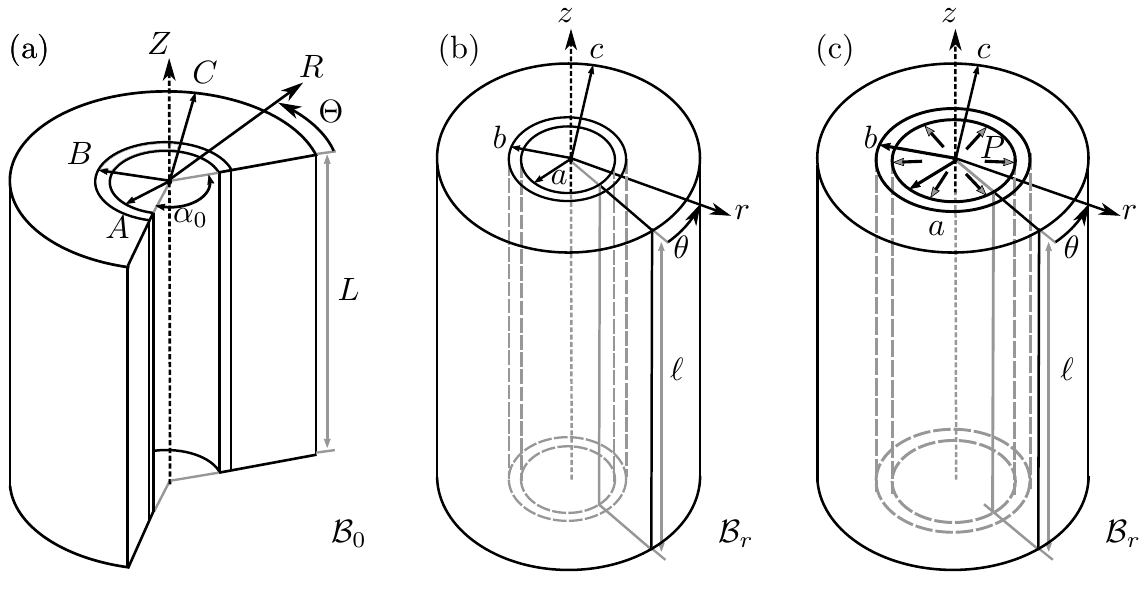}
\caption{The opening angle method: (a) An initially stress-free coated sector is  subject to axial stretch and  bent into (b) a residually-stressed full tube. It can also be subject to (c) an internal pressure. But is that large deformation stable?}
\end{figure}

Consider the  sector of a soft cylindrical tube with geometry delimited in the cylindrical coordinate system $\left\{R,\Theta,Z\right\}$ (and orthonormal basis $\left\{\mathbf{E}_R,\mathbf{E}_\Theta,\mathbf{E}_Z\right\}$) in its natural state $\mathcal{B}_0$ by the  region
\begin{equation}
A\leq R \leq C, \qquad -(2\pi-\alpha_{0})/2 \leq \Theta \leq (2\pi-\alpha_{0})/2, \qquad 0\leq Z \leq L,
\end{equation}
where $A$, $C$ are the radii of the inner and outer faces of the sector, respectively, $L$ is its height, and $\alpha_{0} \in (0,2\pi)$ is the \emph{opening angle}. 
The stress-free circular sector consists of a stiff thin layer placed at the inner side ($A\leq R\leq B$), glued onto a thicker and softer layer located in the outer region $B \leq R \leq C$, where $B$ is the radius of the interface between two layers, as shown on Fig.\ref{FigConf}a. 
From now on, the superscripts $^{(c)}$ and $^{(s)}$ refer to the coating and the substrate, respectively.

The sector is deformed into an intact (circular cylindrical) tube with respect to a cylindrical coordinate system $\left\{r,\theta,z\right\}$ (with orthonormal basis $\left\{\mathbf{e}_r,\mathbf{e}_\theta,\mathbf{e}_z\right\}$) by the following mapping \cite{De10}
\begin{equation}
\label{DefSector}
r=r(R), \qquad \theta=k\Theta,\qquad z=\lambda_{z}Z,
\end{equation}
where 
\begin{equation}
	k=\frac{2\pi}{2\pi-\alpha_{0}}> 1
\end{equation}
is a measure of the opening angle and $\lambda_z \ge1$ is the uniform axial stretch.  
We denote this configuration by $\mathcal{B}_r$ and refer to it as the \emph{residually-stressed configuration}. 
The geometry of the tube is now
\begin{equation}
a\leq r \leq c, \qquad 0 \leq \theta \leq 2\pi, \qquad 0\leq z \leq \ell,
\end{equation}
where $a=r(A)$, $b=r(B)$, $c=r(C)$ and $\ell = \lambda_z L$ is the current tube length as shown on Fig.\ref{FigConf}.

The  associated deformation gradient $\vec{F}$ is
\begin{equation}
\boldsymbol{F} = \dfrac{\text dr}{\text dR}\boldsymbol{e}_r\otimes \boldsymbol{E}_R+\frac{kr}{R}\boldsymbol{e}_\theta\otimes \boldsymbol{E}_\Theta+\lambda_z\boldsymbol{e}_z\otimes \boldsymbol{E}_Z.
\end{equation}
The incompressibility condition,  $\det\mathbf{F}=1$,  and one of the geometric requirements, e.g. $r(A)=a$,  impose
\begin{equation}
r(R)=\sqrt{\frac{R^2-A^2}{k\lambda_z}+a^2}.
\end{equation}

Taking into account the diagonal form of deformation gradient, we introduce the principal stretches
\begin{equation}
\lambda_1=\frac{R}{k\lambda_zr},\qquad \lambda_2=\frac{kr}{R},\qquad\lambda_3=\lambda_z,
\end{equation}
such that $\lambda_1\lambda_2\lambda_3=1$ to satisfy incompressibility.

We take both coating and substrate to be made of isotropic hyperelastic materials with strain energy densities $W^{(c)}$, $W^{(s)}$, respectively, so that the Cauchy stress $\vec \sigma$ is diagonal in the $\vec e_i \otimes \vec e_j$ basis, with components 
\begin{equation}
\label{ConstEq}
\sigma_{rr}^{(l)}=-q^{(l)}+\lambda_1\frac{\partial {W}^{(l)}}{\partial \lambda_1},\quad\sigma_{\theta\theta}^{(l)}=-q^{(l)}+\lambda_2\frac{\partial {W}^{(l)}}{\partial \lambda_2},\quad\sigma_{zz}^{(l)}=-q^{(l)}+\lambda_3\frac{\partial {W}^{(l)}}{\partial \lambda_3}.
\end{equation} 
Here $l=c,s$ and $q^{(l)}$ are the Lagrange multipliers arising from the incompressibility condition.

In the absence of body forces the only non-trivial equation of equilibrium is
\begin{equation}
\label{EqEq}
\dfrac{\partial \sigma_{rr}}{\partial r} ^{(l)}+\frac{\sigma_{rr}^{(l)}-\sigma_{\theta\theta}^{(l)}}{r}=0\quad (l=s,c).
\end{equation}
For the boundary conditions, we assume that the inner (coated) face of the tube at $r=a$ is under internal pressure $P$, that there is perfect bonding between the two layers at the interface $r=b$, and that the outer face at $r=c$ is free of traction:
\begin{equation}
\label{BCs}
\sigma_{rr}^{(c)}(a)=-P,\qquad \sigma_{rr}^{(s)}(b)=\sigma_{rr}^{(c)}(b), \qquad \sigma_{rr}^{(s)}(c)=0.
\end{equation}

By introducing the following quantities \cite{De10},
\begin{equation}
x\equiv k\lambda_{z}\frac{r^{2}}{R^{2}},\qquad x_{a}\equiv k\lambda_{z}\frac{a^{2}}{A^{2}},\qquad x_{b}\equiv k\lambda_{z}\frac{b^{2}}{B^{2}}, \qquad x_{c}\equiv k\lambda_{z}\frac{c^{2}}{C^{2}},
\end{equation}
we may rewrite the principal stretches in terms of $x$ as $\lambda_1=1/\sqrt{k\lambda_zx}$, $\lambda_2=\sqrt{kx/\lambda_z}$ so that the energy density for fixed $\lambda_3=\lambda_z$ may be seen as a function of $x$ only: $\widehat{W}^{(l)}(x) = W^{(l)}(1/\sqrt{k\lambda_zx}, \sqrt{kx/\lambda_z}, \lambda_z)$ for $l=s,c$.

Noting that
\begin{equation}
\sigma_{\theta\theta}^{(l)}-\sigma_{rr}^{(l)}=2x\widehat{W}^{(l)}_{,x}(x)\quad (l=s,c),
\end{equation}
integrating equilibrium equations (\ref{EqEq}) for each layer, and using boundary conditions (\ref{BCs}), we find that the inflating pressure $P$ is
\begin{equation} \label{P}
P=\int^{x_b}_{x_a}\frac{\widehat{W}^{(c)}_{,x}(x)}{1-x}\text dx+\int^{x_c}_{x_b}\frac{\widehat{W}^{(s)}_{,x}(x)}{1-x}\text dx\quad (l=s,c).
\end{equation}
We can also determine the stress components throughout the wall, as
\begin{align} \label{stress-cs}
&\sigma_{rr}^{(s)}(x)=-\int^{x_c}_{x}\frac{\widehat{W}^{(s)}_{,t}(t)}{1-t}\text dt,&& \sigma_{rr}^{(c)}(x)=-\int^{x_b}_{x}\frac{\widehat{W}^{(c)}_{,t}(t)}{1-t}\text dt-\int^{x_c}_{x_b}\frac{\widehat{W}^{(s)}_{,t}(t)}{1-t}\text dt,\nonumber\\[8pt]
&\sigma_{\theta\theta}^{(l)}=\sigma_{rr}^{(l)}+2x\widehat{W}^{(l)}_{,x}(x),&&\sigma_{zz}^{(l)}=\sigma_{rr}^{(l)}+\lambda_3\frac{\partial {W}^{(l)}}{\partial \lambda_3}-\lambda_1\frac{\partial {W}^{(l)}}{\partial \lambda_1}.
\end{align}

For a given geometry of an undeformed coated sector in $\mathcal B_0$, the following quantities are prescribed,
\begin{equation}
\epsilon_B = B^2/A^2 -1, \qquad \epsilon_C = C^2/A^2-1.
\end{equation} 
Then the physics of the stretched and pressurized closed-up cylinder in $\mathcal B_r$ are prescribed by the given strain energy densities $\widehat W^{(l)}$ for coating and substrate, the given axial stretch $\lambda_z$ and the given inner pressure $P$. 
The new geometry is entirely determined by solving the system of three equations for the three unknowns $x_a$, $x_b$, $x_c$ composed by Eq.\eqref{P} and the two relations
\begin{equation}
x_b (\epsilon_B+1) = \epsilon_B + x_a, \qquad x_c(\epsilon_C+1) = \epsilon_C + x_a.
\end{equation}
Then the state of stress is entirely determined by Eqs.\eqref{stress-cs}.

For illustration, in this paper we model the substrate and coating using the Mooney-Rivlin energy density; it reads
\begin{equation} \label{MR}
W^{(l)}=\tfrac{1}{2}C_1^{(l)}\left(\text{tr}(\mathbf C) - 3\right) + \tfrac{1}{2}C_2^{(l)}\left(\text{tr}(\mathbf{C}^{-1})-3\right), \qquad (l=s,c),
\end{equation}
where $C_1^{(l)}>0$ and $C_2^{(l)}>0$ are material constants and $\mathbf C = \mathbf F^T \mathbf F$ is the right Cauchy-Green deformation tensor. 
This model is quite general because it recovers, at the same level of approximation \cite{DeGM10}, the most general model of isotropic, incompressible, third-order weakly non-linear elasticity,
\begin{equation}
W = \mu\, \text{tr}(\mathbf E^2) + \tfrac{1}{3}A \, \text{tr}(\mathbf E^3),
\end{equation}
where $\mathbf E = 2\mathbf C + \mathbf I$ is the Green-Lagrange strain tensor, $\mu$ is the Lam\'e coefficient of linear elasticity, and $A$ is the Landau coefficient of third-order elasticity
(The connections between the constants are
$\mu = C_1+C_2$,  $A =  -4C_1 - 8C_2$.)
For the Mooney-Rivlin material \eqref{MR}, we
have
\begin{equation}
\widehat W^{(l)}(x) =\tfrac{1}{2}(C_1^{(l)} + C_2^{(l)}\lambda_z^2)\left( \dfrac{kx}{\lambda_z} + \dfrac{1}{k\lambda_z x}\right) + \text{constant},
\end{equation}
which provides explicit expressions for the stress components in Eq.\eqref{stress-cs}.
Hence
\begin{equation}
\sigma^{(s)}_{rr} = \dfrac{C_1^{(s)} \lambda_z^{-1} + C_2^{(s)} \lambda_z}{2k}\left[(1-k^2)\ln \left(\dfrac{x-1}{x_c-1}\right) - \ln \left(\dfrac{x}{x_c}\right) + \dfrac{1}{x} - \dfrac{1}{x_c}\right],
\end{equation}
and so on for the other components.

\begin{figure}[!htb]
\includegraphics[width=0.9\linewidth,grid]{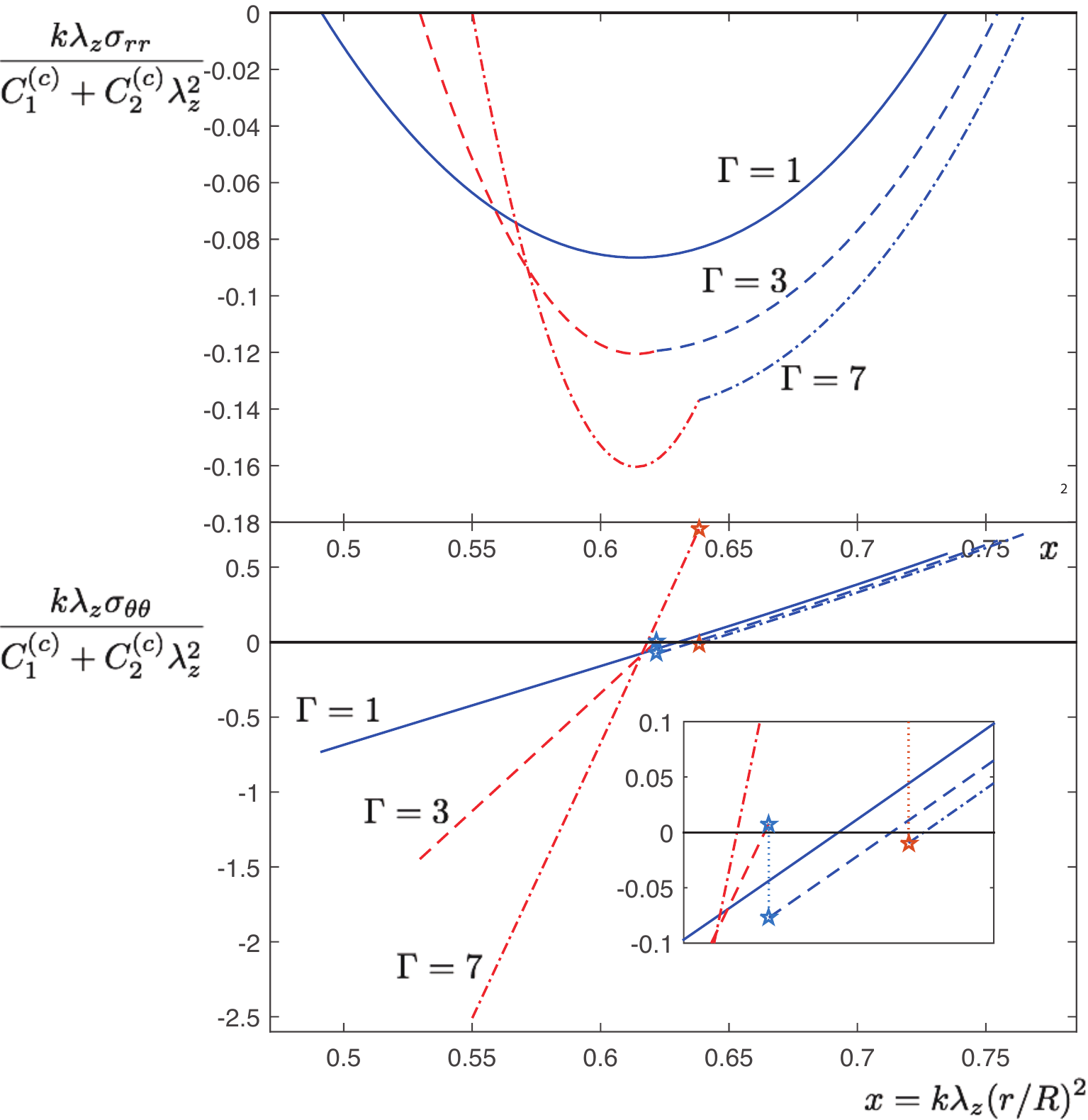}
\caption{Non-dimensional radial $\sigma_{rr}$ and circumferential $\sigma_{\theta\theta}$ stresses through two-layered wall of coating (red) and substrate (blue) modeled as Mooney-Rivlin materials with corresponding material constants $C_i^{(c)}$ and $C_i^{(s)}$ related by $C_i^{(c)}=\Gamma C_i^{(s)}$ ($i=1,2$, $j=0,1,2$), where $\Gamma \ge 1$ is the stiffness contrast between the coating and the substrate.}
\label{large}
\end{figure}

For an example, assume that the coating is  $\Gamma$ times stiffer than the substrate, in the sense that $C_1^{(c)} = \Gamma C_1^{(s)}$, $C_2^{(c)} = \Gamma C_2^{(s)}$, where $\Gamma \ge 1$ is the stiffness contrast factor. 
Then we consider how the stresses are distributed along the radial axis for different stiffness factors $\Gamma$. 
We take the case where there is no inner pressure ($P=0$) and the opening angle is $139^\circ$. 
In the undeformed geometry we take $A=13$mm, $B=14.5$mm, $C=18$mm. 
Fig.\ref{large} illustrates the distribution of stresses along the thickness of the wall of closed-up cylinders, for a uniform material $(\Gamma=1.0$), and for two-layered solids with moderately ($\Gamma=3.0$) and significantly ($\Gamma=7.0$) stiffer coatings compared to substrates. 
We clearly observe the jump in the circumferential stresses at the interface between coating and substrate, as expected. 

\section{Wrinkling of a coated sector}
\label{section3}


Here we study the stability of a coated sector closed into a pressurized cylinder. 
We signal the onset of instability by the existence of small-amplitude wrinkles, solutions to the incremental equations of equilibrium. 
From experimental observations, we know that they should be varying sinusoidally along the circumference of the tube, with amplitude decay from the inner face to the outer face. 
The analysis for the existence of such wrinkles can be put together from the results of the previous section and those of Destrade et al. \cite{De10} and we omit the details to save space.

In short, the wrinkles exist when the following boundary value problem is solved for $\mathbf z^{(l)} = \mathbf z^{(l)}(x)$, ($l=s,c$), the $2 \times 2$ Hermitian \emph{surface impedance matrix} \cite{DeAC09}. 
\begin{itemize}
\item[\emph{(i)}] Initial condition: $\mathbf z^{(s)}(x_{c})=\mathbf 0$; 
\item[\emph{(ii)}] Numerical integration of the differential Riccati matrix equation
\begin{equation} \label{riccati}
\dfrac{\text d}{\text dx} \mathbf z^{(l)}= \dfrac{1}{2x(1-x)}\left[ \mathbf z^{(l)}\mathbf G_2^{(l)} \mathbf z^{(l)} + \text i \left(\mathbf G_1^{(l)}\right)^\dagger \mathbf z^{(l)} -  \text i \mathbf z^{(l)} \mathbf G_1^{(l)} + \mathbf G_3^{(l)}\right],
\end{equation}
in the substrate ($l=s$), from $x_c$ to $x_b$;
\item[\emph{(iii)}] Interfacial condition: $\mathbf z^{(c)}(x_b) = \mathbf z^{(s)}(x_b)$;
\item[\emph{(iv)}]  Numerical integration of the differential Riccati matrix equation \eqref{riccati} in the coating ($l=c$), from $x_b$ to $x_a$;
and 
\item[\emph{(v)}] Target condition: 

\begin{equation}
\det \left(\mathbf z^{(c)}(x_a) + P \begin{bmatrix} 1 & \ii n \\ - \ii n & 1 \end{bmatrix}\right) = 0.
\label{target}
\end{equation}

\end{itemize}

In Eq.\eqref{riccati}, $\dagger$ denotes the Hermitian transpose and the Stroh sub-matrices $\mathbf G_i$ have components \cite{De10},
\begin{equation}
\boldsymbol{G}_1 = 
\begin{bmatrix}
\text i & -n\\
-n(1-\sigma) & - \text i(1 - \sigma)
\end{bmatrix}, \quad
\boldsymbol{G}_2 = 
\begin{bmatrix}
0&0\\
0&1/\alpha
\end{bmatrix}, \quad
\boldsymbol{G}_3 = 
\begin{bmatrix}
\kappa_{11} & \text i \kappa_{12}\\
- \text i\kappa_{12} & \text i \kappa_{22}
\end{bmatrix},
\end{equation}
where the superscript ``${(l)}$'' is understood, $n$ denotes the wrinkling mode (number of wrinkles in the circumference), and
\begin{align}
& \kappa_{11} = 2 \beta + 2 \alpha(1 - \sigma) + n^2[\gamma - \alpha(1-\sigma)^2],\notag \\
& \kappa_{12} = n(2\beta + \gamma + \alpha (1 - \sigma^2),\notag \\
& \kappa_{22} = \gamma - \alpha (1 - \sigma)^2 + 2 n^2(\beta + \alpha (1 - \sigma).
\end{align}
Here, in general,
\begin{equation}
\alpha = \frac{2x\widehat{W}_{,x}(x)}{k^2 x^2 - 1},\quad
\gamma = k^2 x^2 \alpha, \quad 
\beta = 2x^2 \widehat{W}_{,xx}(x) + x \widehat{W}_{,x}(x) - \alpha,
\quad
\sigma = \sigma_{rr}/\alpha,
\end{equation}
and in particular for the Mooney-Rivlin model,
\begin{equation}
\alpha = (C_1\lambda_z^{-1} + C_2\lambda_z)\dfrac{1}{k x}, \qquad
\gamma =  (C_1\lambda_z^{-1} + C_2\lambda_z)k x, \qquad \beta = \tfrac{1}{2}(\alpha + \gamma).
\end{equation}

Finally, the derivation of the target condition \eqref{target}  is detailed in the appendix.


\section{Experimental \& numerical results}
\label{section4}


Here we implement the stability analysis described in the previous section for two cases: polymers and biological tissues. The algorithm is illustrated in Fig.\ref{stability}(a). Essentially, we implement the steps \textit{(i)}-\textit{(iv)} and iterate over $\alpha_0$ until the target condition \textit{(v)} is reached. 
We denote by $\alpha_{\mathrm{cr}}=\alpha_0$  the critical opening angle at which wrinkles form when the sector is closed into an intact tube, i.e., the value of $\alpha_0$ when the target condition is reached.


\subsection{Results for polymers}
\label{Results for polymers}

\begin{figure}[ht!]
\centering
\includegraphics[width=1.0\linewidth,grid]{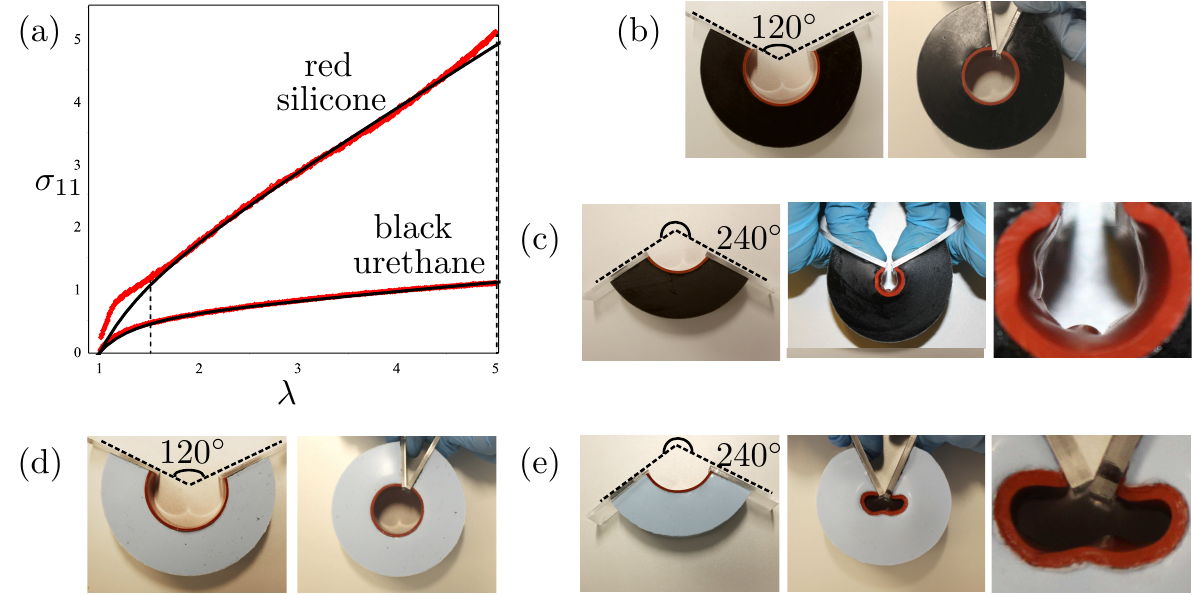}
\caption{(a) Tensile tests for red silicone and black urethane. The early part of the data for silicone was discarded as unreliable and the curve-fitting to the Mooney-Rivlin models was done over the $1.5 \le \lambda \le 5.0$ range indicated by the dashed lines, yielding a relative error of less than 5\%.
(b) Sector with opening angle $120^\circ$, black urethane substrate and red silicone coating. No wrinkles form when the sector is closed into an intact tube. (c) Sector with opening angle $240^\circ$, black urethane substrate and red silicone coating. Six wrinkles form shortly before the sector is closed into an intact tube. (d), (e) Similar results for sectors with white silicone substrate and red silicone coating, and opening angles $120^\circ$ and $240^\circ$, respectively. }
\label{exps}
\end{figure}

For our first experiment, we used artificial materials, namely relatively stiff silicone (red) for the coating, urethane (black) and very soft silicone (white) for the substrate. 
We subjected each material to a tensile test using a MTS electromechanical material characterization machine. We then determined the Mooney-Rivlin constants by curve-fitting over a useable range of data, and found that $C_1^{(c)} = 0.98$, $C_2^{(c)}=0.021$ (MPa) for the red silicone, and $C_1^{(s)} = 0.14$, $C_2^{(s)}=0.41$ (MPa) for the black urethane,  see Fig.\ref{exps}(a). 
We then glued a 1.6mm thick red silicone layer onto a 26.9mm thick black urethane sector ($B=23.93$mm, $C=50.83$mm) and produced two coated sectors, one with opening angle $120^\circ$, the other with opening angle $240^\circ$, see Fig.\ref{exps}(b) and (c). We produced similar sectors using white urethane as the substrate, see Fig.\ref{exps}(d).

We found that, for the black urethane substrate, no wrinkles formed when the former sector was closed (Fig.\ref{exps}(b)), while for the latter sector six wrinkles formed shortly before the sector became intact (Fig.\ref{exps}(c)). 
Thus we would expect the critical opening angle at which wrinkles form when the sector becomes intact to be somewhere between $120^\circ$ and $240^\circ$. 
To check this assertion, we performed the stability analysis described in the previous section for the same dimensions and material parameters as in the experiments. We found that the critical opening angle was $209^\circ$ with corresponding mode number $n=4$, which supports our previous hypothesis.

\subsection{Results for soft tissues}  


Here we perform the stability analysis using the dimensions and material parameters  which are of the same order of magnitude as those  of a rabbit carotid artery, as collected by Holzapfel et al. \cite{holz}. 

The artery consists of three layers: the intima, the media and the adventitia. 
However, the intima is very thin and not very stiff (at least in healthy young individuals),  and so we can use our two-layer model  with the dimensions  \cite{holz} $B-A=0.26$mm, $C-B=0.12$mm, $A=1.43$mm, along with an axial stretch $\lambda_z=1.695$. 

For the material parameters,  Holzapfel et al. \cite{holz} used an anisotropic model. Here we have only considered isotropic models, and so we set to zero Holzapfel et al.'s anisotropic parameters to make a (somewhat arbitrary) connection with their measurements. 
Moreover, Holzapfel et al. \cite{holz} did not consider a dependence of $W$ on the second invariant of strain $\text{tr}(\mathbf C^{-1})$, so here we take $C_2^{(c)}=0$, $C_2^{(s)}=0$. 
For the other (neo-Hookean) parameters, we have $C_1^{(c)}=3$kPa, $C_1^{(s)}=0.3$kPa, in line with Holzapfel et al.'s \cite{holz} values of the shear modulus for the artery's elastin matrix.

We perform the stability analysis over a physiological pressure range  \cite{dominguez} of 0-170 mmHg. We plot the results in Fig.\ref{stability}(c) for the non-dimensional measure of pressure $\hat{P}=P/C_1^{(s)}$. 
	
Then the  physiological pressure range corresponds to $0 \le \hat P \le 75.5$. 

\begin{figure}[ht!]
\centering
\includegraphics[width=\textwidth]{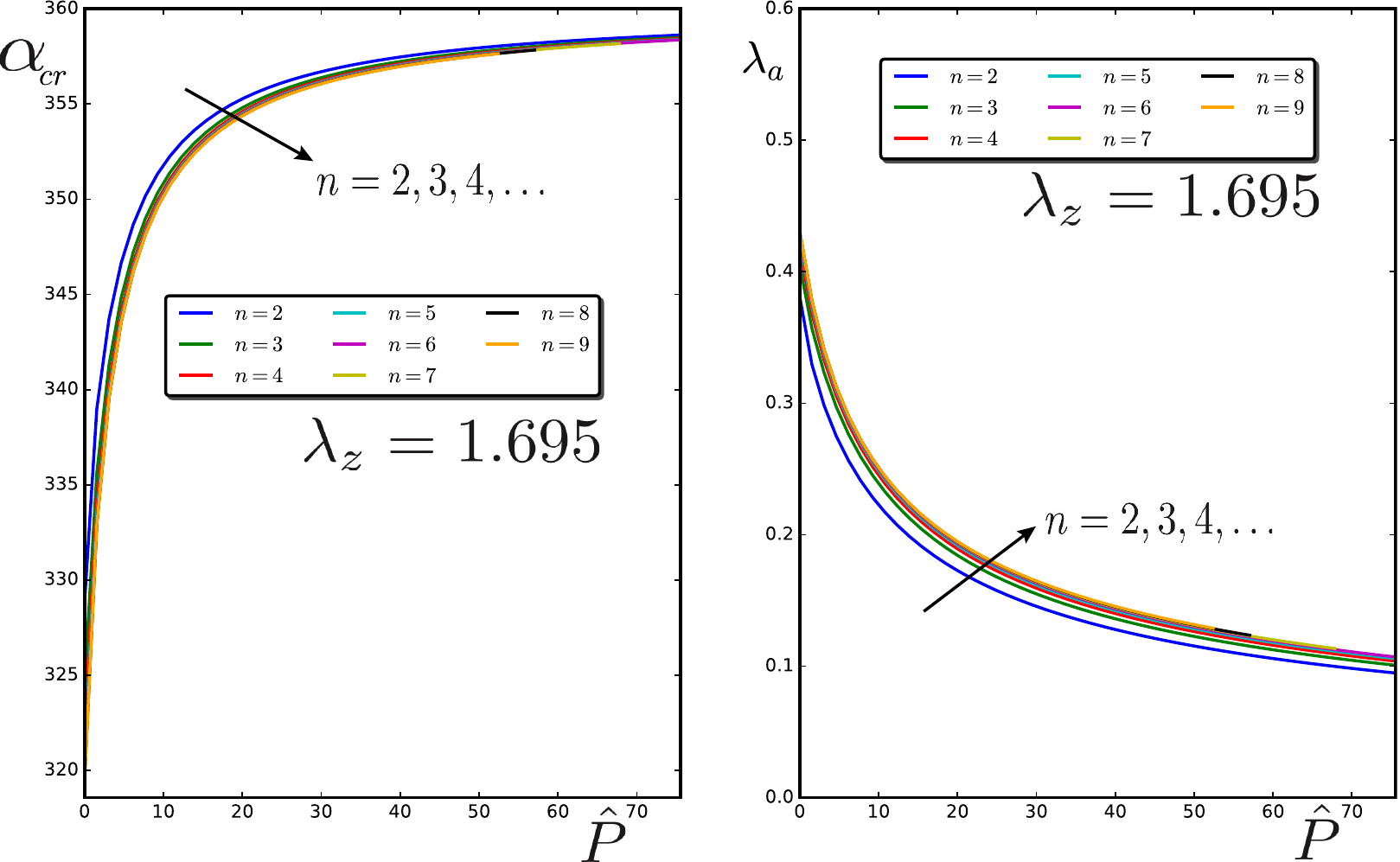}
\caption{ (a) Plots of the critical opening angle $\alpha_\text{cr}$ for several mode numbers $n$ versus the non-dimensional pressure $\hat{P}$ using the material parameters and dimensions comparable with those of a rabbit artery \cite{holz}, when it is subject to an axial stretch $\lambda_z = 1.695$.
(b) Plots of the critical circumferential stretch $\lambda_a$ on the inner face of the intact tube at buckling versus the pressure $\hat P$.}
\label{fig:stretch}
\end{figure}

First we plot the curves giving the critical opening angle $\alpha_\text{cr}$ against the pressure $\hat P$ for increasing values of the mode number $n = 2,3,4,\ldots$.
Each curve is a bifurcation plot: at a given pressure $\hat P$, a tube with opening angle larger than $\alpha_\text{cr}$ will buckle when it is bent into an intact closed tube; in order not to buckle, a sector must have an opening angle which is less than the smallest critical angle from all curves. 
Here we find that all curves for mode numbers $n \ge 5$ are all below those for $n=2,3,4$ and are virtually indistinguishable one from another, see Fig.\ref{fig:stretch}.
Hence our analysis does not allow us to determine the mode number precisely here, in contrast to the scenario of Section \ref{Results for polymers}.

From the plots we see that when there is no internal pressure ($\hat P=0$), only sectors with an opening angle greater than $\alpha_\text{cr} \simeq 320^\circ$ will buckle when closed into an intact tube. 
This value is significantly above the recorded opening angle for the rabbit artery \cite{holz}, which was $160^\circ$. 
Hence we would expect (provided the crudeness of our modelling arteries here is overlooked) that the rabbit artery is smooth when it is not subject to internal pressure.

We also observe that as the internal pressure increases, the critical opening angle increases, with asymptotic behaviour  $\alpha_\text{cr} \to 360^\circ$  as $\hat P \to \infty$.
Hence buckling can be eliminated by applying an internal pressure, which is in line with our intuition and with, for example, experiments on a rat's pulmonary artery \cite{fung}, see Fig.\ref{stability}(b).

\begin{figure}[ht!]
\centering
\includegraphics[width=\textwidth]{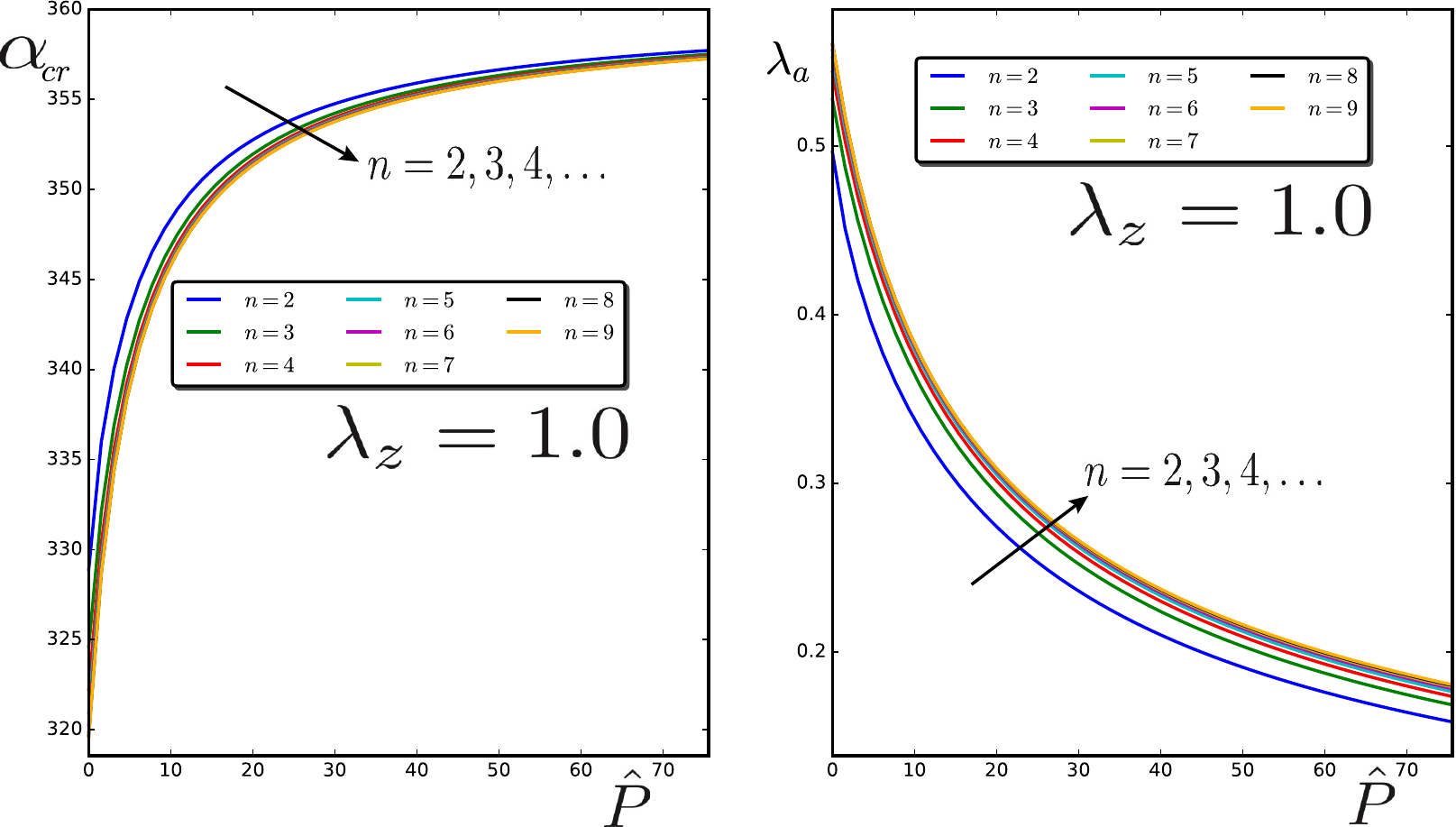}
\caption{ (a) Plots of the critical opening angle $\alpha_\text{cr}$ for several mode numbers $n$ versus the non-dimensional pressure $\hat{P}$ using the material parameters and dimensions comparable with those of a rabbit artery \cite{holz}, when it is not subject to an axial stretch ($\lambda_z = 1.0$).
(b) Plots of the critical circumferential stretch $\lambda_a$ on the inner face of the intact tube at buckling versus the pressure $\hat P$.}
\label{fig:no-stretch}
\end{figure}

For comparison, we also plot the curves obtained in the case of no axial stretch, $\lambda_z =1$, see Fig.\ref{fig:no-stretch}. 
We find that the axial stretch makes the sector more stable with respect to bending into an intact tube (the values of $\alpha_\text{cr}$ are higher when $\lambda_z>1$ than when $\lambda_z=1$).
To complete the picture, we also provide the plots of the variations of the critical circumferential stretch $\lambda_a$ (contraction stretch on the inner face of the intact tube). 

\section{Discussion}

\begin{figure}[ht!]
\centering
\includegraphics[width=0.88\linewidth,grid]{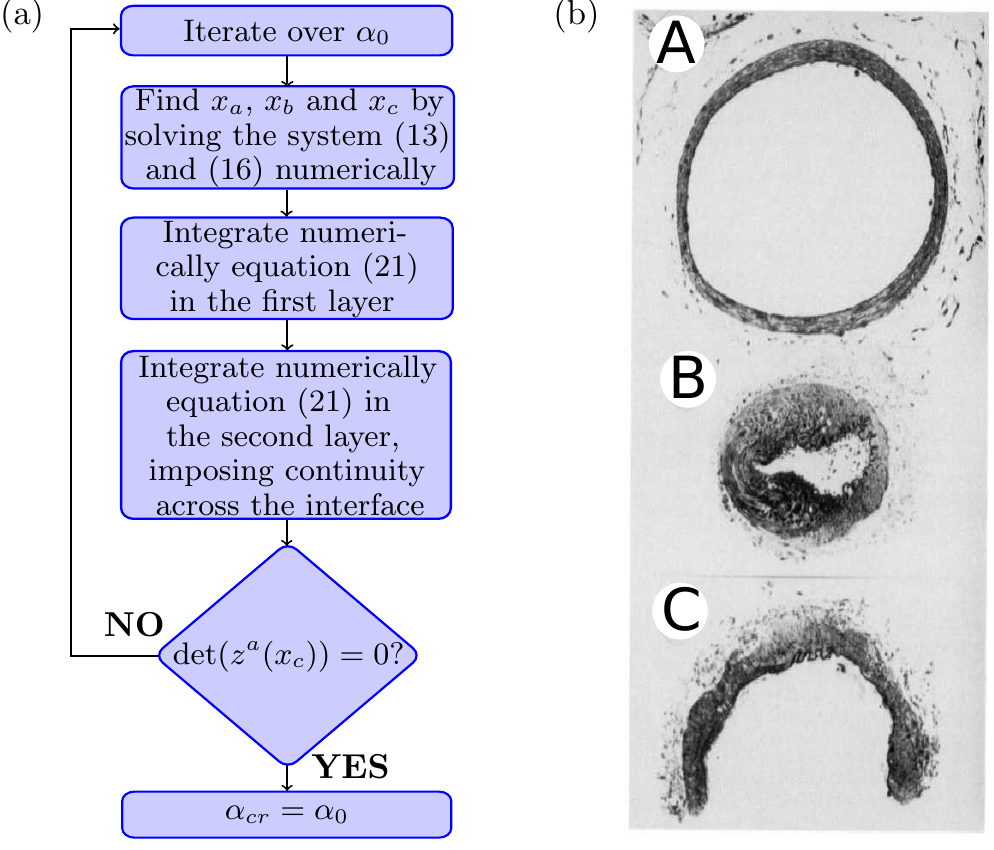}
\caption{ (a) Flow chart illustrating the algorithm used to find the critical opening $\alpha_{\mathrm{cr}}$ for given $P$ and $n$. (b) Rat pulmonary artery at three different states:  (A) Intact with low internal pressure of 15 mmHg and smooth internal surface; (B) Intact with no pressure and buckled internal face; (C)  Cut open (image retrieved from \cite{fung}). (c) Plot of the critical opening angle (for mode $n=4$) versus the non-dimensional pressure $\hat{P}$ using the material parameters and dimensions of a rabbit artery \cite{holz}. Solid line: axial stretch $\lambda=1.695$, dashed line: $\lambda=1$. }
\label{stability}
\end{figure}

Often it is assumed that a stable deformation of a sector into an intact tube exists. These ``opening angle'' deformations are then used to estimate the residual stresses in the material \cite{herrera}. Here we have shown that, depending on the material properties and dimensions, wrinkling may occur before the sector becomes intact, which would be followed by further buckling and creases when the sector is closed. 
Our results have important implications for finite element reconstructions of the opening angle method.
First, a stiffer coating will lead to instabilities in finite element simulations, earlier than for a homogeneous sector \cite{De10, herrera}. Second, if the wrinkles occur, then our analysis is a first step towards providing meaningful precursors to creases (see Fig.\ref{residual} and Fig.\ref{stability}(b)).

We  also showed that wrinkles  can be eliminated by applying an internal pressure, as has been confirmed in experiments.

Our method could also be applied to other tissues such as the esophagus, which is often modeled as a two-layered structured, and in which wrinkles and creases have been observed \cite{sokolis}. 
However, it is important to consider the limitations of  our model.  For example, in the iliac artery of an 81 year old human, buckling of the intima in the zero-pressure state leading to delamination has been observed \cite{holz2}. As has been noted, the intima, one of the three layers of the artery, becomes thicker and stiffer with age. Evidently, there are residuals stresses present leading to buckling, but clearly a three-layer model would be necessary to investigate such an occurrence. Furthermore, each layer of the artery is highly anisotropic due to the presence of collagen fibers \cite{holz}, and so a more realistic model would reflect this fact.


\section*{Acknowledgments}


MD and  RM are grateful to the Irish Research Council for support through a Government of Ireland Postgraduate Scholarship.
MD and IL thank the NUI Galway College of Science for support with the Summer Internship Program. TS is grateful to Aleksander Czekanski and IDEA-Lab research group from York University for their valuable support in performing the experiments.
	
Finally, we are grateful to Valentina Balbi (Galway) for helpful discussions on the incremental problem.


\section*{Appendix: Derivation of the target condition \eqref{target}}


At the coating/vacuum interface, the incremental nominal traction is \cite{DeOO07}
\begin{equation} \label{interf}
\mathbf s^T \mathbf e_r = \left[\boldsymbol \sigma^* + P(\text{grad}\ \mathbf u)^T\right]\mathbf e_r,
\end{equation}
where $\mathbf u$ is the incremental mechanical displacement, and $\boldsymbol \sigma^*$ is the Cauchy incremental stress in the $0 \le r \le a$ region. 
But that space is under constant hydrostatic pressure $P$ and has no constitutive law to speak of, being the vacuum, so that $\boldsymbol \sigma^* \equiv \mathbf 0$.
Also, the displacement gradient has components \cite{De10}
\begin{equation}
\text{grad}\ \mathbf u = \begin{bmatrix} \dfrac{\partial u}{\partial r} & \dfrac{1}{r}\left(\dfrac{\partial u}{\partial \theta} - v\right) \\[12pt] 
 \dfrac{\partial v}{\partial r} & \dfrac{1}{r} \left(u + \dfrac{\partial v}{\partial \theta} \right) \end{bmatrix},
 \end{equation}
 in the $\mathbf e_i \otimes \mathbf e_j$ basis. 

For displacements of the form
\begin{equation}
\{ u , v \} = \{ U(r) \ee^{\ii n\theta}, V(r) \ee^{\ii n\theta} \},
\end{equation}
describing prismatic wrinkles, the incremental nominal traction is also of a similar form:
\begin{equation}
\{ s_{rr} , s_{r\theta} \} = \{ S_{rr}(r) \ee^{\ii n\theta}, S_{r\theta}(r) \ee^{\ii n\theta} \},
\end{equation}
where $U$, $V$, $S_{rr}$, $S_{r\theta}$ are functions of $r$ only. 
Then \eqref{interf} reads
\begin{equation} \label{interf2}
r\begin{bmatrix} S_{rr}\\S_{r\theta}\end{bmatrix} = P\begin{bmatrix} rU'\\ \ii n U - V \end{bmatrix} 
= P\begin{bmatrix} -U - \ii nV\\ \ii n U - V \end{bmatrix},
\end{equation}
at $r=a$, where for the second equality we used the incremental incompressibility equation,
\begin{equation}
\text{div}\ \mathbf u = \dfrac{\partial u}{\partial r} + \dfrac{1}{r} \left(u + \dfrac{\partial v}{\partial \theta} \right) = 
(rU' + U + \ii nV)\dfrac{\ee^{\ii n\theta}}{r} = 0.
\end{equation}

On the other hand, the traction is related to the displacement by the surface impedance matrix \cite{DeAC09}:
\begin{equation}
r\begin{bmatrix} S_{rr}\\S_{r\theta}\end{bmatrix} = \mathbf z^{(c)} \begin{bmatrix} U\\V\end{bmatrix}.
\end{equation}
In particular, at the $r=a$ interface, we have by \eqref{interf2}
\begin{equation}
P\begin{bmatrix} -U(a) - \ii nV(a) \\ \ii n U(a) - V(a) \end{bmatrix} =  \mathbf z^{(c)}(a)\begin{bmatrix} U(a)\\ V(a) \end{bmatrix},
\end{equation}
from which the target condition \eqref{target} follows
(see Balbi and Ciarletta \cite{BaCi15} for an early, but not entirely correct, derivation of the target condition).

\section*{References}
\vspace{12pt}




\begin{thebibliography}{99}


\bibliographystyle{elsarticle-num.bst}


\bibitem{BaCi15}
Balbi, V., \& Ciarletta, P. (2015). Helical buckling of thick-walled, pre-stressed, cylindrical tubes under a finite torsion. Mathematics and Mechanics of Solids, 20, 625-642. https://doi.org/10.1177/1081286514550570
	
\bibitem{DeAC09}
Destrade, M., Annaidh, A. N., \& Coman, C. D. (2009). Bending instabilities of soft biological tissues. International Journal of Solids and Structures, 46, 4322-4330.

\bibitem{DeGM10}
Destrade, M., Gilchrist, M. D., \& Murphy, J. G. (2010). Onset of no-linearity in the elastic bending of blocks. Journal of Applied Mechanics, 77, 061015.

\bibitem{De10}
Destrade, M., Murphy, J. G., \& Ogden, R. W. (2010). On deforming a sector of a circular cylindrical tube into an intact tube: existence, uniqueness, and stability. International Journal of Engineering Science, 48, 1212-1224.

\bibitem{dominguez}
Dominguez, R. (1927). The systolic blood pressure of the normal rabbit measured by a slightly modified van Leersum method, 46, 443-461.

\bibitem{fung}
Fung, Y. C., \& Liu, S. Q. (1992). Strain distribution in small blood vessels with zero-stress state taken into consideration. The American journal of Physiology, 262, H544-52.

\bibitem{herrera}
Garcia-Herrera, C. M., Bustos, C. A., Celentano, D. J., \& Ortega, R. (2016). Mechanical analysis of the ring opening test applied to human ascending aortas. Computer Methods in Biomechanics and Biomedical Engineering, 1-11.

\bibitem{holz}
Holzapfel, G. A., Gasser, T. C., \& Ogden, R. W. (2000). A new constitutive framework for arterial wall mechanics and a comparative study of material models. Journal of Elasticity and the Physical Science of Solids, 61, 1-48.

\bibitem{holz2}
Holzapfel, G. A., Sommer, G., Gasser, C. T., \& Regitnig, P. (2005). Determination of layer-specific mechanical properties of human coronary arteries with nonatherosclerotic intimal thickening and related constitutive modeling. American Journal of Physiology-Heart and Circulatory Physiology, 289, H2048-H2058.

\bibitem{OmFu90}
Omens, J. H., \& Fung, Y. C. (1990). Residual strain in rat left ventricle. Circulation Research, 66, 37-45.

\bibitem{DeOO07}
Ott\'enio, M., Destrade, M., \& Ogden, R. W. (2007). 
Acoustic waves at the interface of a pre-stressed incompressible elastic solid and a viscous fluid. 
International Journal of Non-Linear Mechanics, 42, 310-320. https://doi.org/10.1016/j.ijnonlinmec.2006.10.001

\bibitem{sokolis}
Sokolis, D. P. (2010). Strain-energy function and three-dimensional stress distribution in esophageal biomechanics. Journal of Biomechanics, 43, 2753-2764.


\end{thebibliography}
\end{document}